\documentclass [12pt]{article}

\begin{document}
\begin{center}
\textbf{\huge Canonical Entropy of charged black hole}
\end{center}

\begin{center}
Zhao Ren\footnote{Corresponding address: Department of Physics,
Shanxi Datong University, Datong 037009, P.R.China; E-mail
address: zhaoren2969@yahoo.com.cn }

Department of Physics, Shanxi Datong University, Datong 037009
P.R.China\\ Department of Applied Physics, Xi'an Jiaotong
University, Xi'an 710049 P.R.China

Li Huai-Fan \\
Department of Physics, Shanxi Datong University, Datong 037009
P.R.China

Zhang Sheng-Li\\
Department of Applied Physics, Xi'an Jiaotong University, Xi'an
710049 P.R.China

\end{center}

\begin{center}
\textbf{Abtract}
\end{center}

Recently, Hawking radiation of the black hole has been studied by
using the tunnel effect method. It is found that the radiation
spectrum of the black hole is not a strictly pure thermal
spectrum. How does the departure from pure thermal spectrum affect
the entropy? This is a very interesting problem. In this paper, we
calculate the partition function through energy spectrum obtained
by using the tunnel effect. From the relation between the
partition function and canonical entropy, we can derive the
entropy of charged black hole. In our calculation, we consider not
only the correction to the black hole entropy due to fluctuation
of energy, but also the effect of the change in the black hole
charges on entropy. There is not any assumption. This makes our
result more
reliable.\\
PACS numbers: 04.70.Dy, 04.60.Pp
\vspace{0.5cm}\\
\textbf{\large 1. Introduction}

Hawking [1] interpreted the quantum effect of the black hole as emission of
thermal radiant spectrum from event horizon, which sets a milestone in black
hole physics. The discovery of this effect not only solved the problem in
black hole thermodynamics, but also revealed the relation among quantum
mechanics, thermodynamics and gravitation. Studying the thermal properties
of various black holes is one of the important subjects of black hole
physics. Hawking pointed out that vacuum fluctuation near the surface of the
black hole would produce virtual particle pair. When the virtual particles
with negative energy come into black hole via tunnel effect, the energy of
the black hole will decrease. At the same time, the particle with positive
energy may thread out the gravitational region outside the black hole. This
process is equivalent to the emission of a particle from the black hole.
However, in Hawking's proof there is not any potential barrier in the
tunnel.

Parikh and Wilczek [2] discussed Hawking radiation by tunnel
effect. They thought that tunnels in the process of the particle
radiation had no potential barrier before particles radiated.
Potential barrier is produced by radiation particles itself. That
is, during the process of tunnel effect creation, the energy of
the black hole decreases and the radius of the black hole horizon
reduces. The horizon radies gets a new value that is smaller than
the original value. The decrease of radius is determined by the
value of energy of radiation particles. There is a classical
forbidden band-- potential barrier between original radius and the
one after the black hole radiates. Parikh and Wilczek skillfully
obtained the radiation spectrum of Schwarzschild and
Reissner-Nordstrom black holes. Refs.[3-14] developed the method
proposed by Parikh and Wilczek. They derived the radiation
spectrum of the black hole in all kinds of space-time.
Refs.[13-16] obtained radiation spectrum of Hawking radiation
after considering the generalized uncertainty relation. And
Angheben, Nadalini, Vanzo and Zerbini have computed the radiation
spectrum of the arbitrary dimensional black hole ,but haven't
obtained universal expression for radiation spectrum of static
mass equal to zero and charged particle.

In this paper, firstly we obtain the radiation spectrum of the
black hole which is independent of whether static mass of
radiation particle equal to zero by quantum statistical method,
and then using this radiation spectrum, we calculate the black
hole canonical entropy and derive the canonical entropy of the
charged black hole.\vspace{0.5cm}\\
\textbf{\large 2. Canonical partition function of charged black
hole}

For static or stable space-times, since the metric is independent
of time variable, in the spatial region we can constitute a
contemporaneous plane that surrounds the black hole. We put the
black hole in contact with a thermal radiant field with
temperature $T(T$ is the radiation temperature of the black hole).
We require that $R > > r_H $, where $R$ is the radius of the
contemporaneous plane and $r_H $ is the horizon radius of the
black hole. Since the radius of the contemporaneous plane is much
bigger than the horizon radius of the black hole, we can consider
the region surrounded by this contemporaneous plane as an isolated
thermodynamic system with conserved energy. This region can be
divided into three parts: the naked black hole, horizon surface
and radiation field. Suppose that the total energy is $E^0$, the
total charges is $Q^0$, the initial energy of the naked black hole
is $E$, which is Arnowitt-Deser-Misner (ADM) mass, and the charge
is $Q$. Initial energy of the horizon surface of the black hole
and the charge are zero. The energy of the radiation field is $E_r
$, and the charge is $Q_r $. We know that the location of the
black hole horizon $r_H $ is a function of energy and charge. It
is denoted as $r_H (E,Q)$. When the black hole has Hawking
radiation, the horizon location changes. Suppose that the energy
of Hawking radiation particles is $E_s $ and the electric charge
is $Q_n $, then the horizon location will change from $r_H (E,Q)$
to $r_H (E - E_s ,Q - Q_n )$. At this time, a quantum energy layer
with energy $E_s $ and the electric charge $Q_n $ is formed
between the two horizons. Because the black hole and the horizon
are put in contact with a thermal radiant field with temperature
$T$, the temperature is invariant during the creation process of
radiation. Therefore, we can assume that in this process the
temperature of the black hole is a constant. This hypothesis is
consistent with the one used when the Hawking radiation is
discussed via tunnel effect. At this time, there is no energy
exchange between energy layer and the radiation field. As a
result, the energy of energy layer and the energy of naked black
hole are conservative, and the charge of energy layer and the
charge of naked black hole are conservative.

When the quantum energy layer is at state $s$ with the charge $Q_n
$ and energy $E_s $, the naked black hole can be at any
microscopic state with the charge $Q - Q_n $ and energy $E - E_s
$. Let $\Omega (E - E_s ,Q - Q_n )$ denote the number of
microscopic state of the naked black hole with the charge $q = Q -
Q_n $ and energy $E_b = E - E_s $. When the quantum energy layer
is at state $s$, the number of microscopic states of the naked
black hole and quantum energy layer is $\Omega (E - E_s ,Q - Q_n
)$. According to the equiprobable principle, every microscopic
state of the compound system that consists of the naked black hole
and quantum energy layer appears with the same probability. The
probability that quantum energy layer is at state $s$ is
proportional to $\Omega (E - E_s ,Q - Q_n )$. That is

\begin{equation}
\label{eq1} \rho _{s,n} \propto \Omega (E - E_s ,Q - Q_n ).
\end{equation}

Because the number of microscopic state of the system is very big,
in convenience, we discuss $\ln \Omega $. We expand $\ln \Omega $
as a power series with respect to $E_s $ and $Q_n$. We have

\[
\ln \Omega (E - E_s ,Q - Q_n ) = \ln \Omega (E,Q) \] \[+ \left(
{\frac{\partial \ln \Omega }{\partial E_b }} \right)_{E_s = 0,Q_n
= 0} ( - E_s ) + \frac{1}{2}\left( {\frac{\partial ^2\ln \Omega
}{\partial E_b^2 }} \right)_{E_s = 0,Q_n = 0} E_s^2 + \cdots
\]

\begin{equation}
\label{eq2}
 + \left( {\frac{\partial \ln \Omega }{\partial q}} \right)_{E_s = 0,Q_n
= 0} ( - Q_n ) + \frac{1}{2}\left( {\frac{\partial ^2\ln \Omega }{\partial
q^2}} \right)_{E_s = 0,Q_n = 0} Q_n^2 + \cdots .
\end{equation}
The first term in the right hand side of Eq.(\ref{eq2}) is a
constant. So (\ref{eq1}) can be rewritten as

\begin{equation}
\label{eq3}
\rho _{sn} \propto \exp [ - \beta E_s + \beta _2 E_s^2 - \alpha Q_n + \alpha
_2 Q_n^2 + \cdots ],
\end{equation}

\noindent
where $\alpha _k = \frac{1}{k!}\left( {\frac{\partial ^k\ln \Omega
}{\partial q^k}} \right)_{E_s = 0,Q_n = 0} $,$\beta _k = \frac{1}{k!}\left(
{\frac{\partial ^k\ln \Omega }{\partial E_b^k }} \right)_{E_s = 0,Q_n = 0}
$. In statistical physics, logarithm of the number of microscopic state of
the system should be the entropy of the system. That is

\begin{equation}
\label{eq4}
S = \ln \Omega .
\end{equation}
Eq.(\ref{eq2}) can be rewritten as

\begin{equation}
\label{eq5}
S(E - E_s ,Q - Q_n ) - S(E,Q) = - \alpha Q_n + \alpha _2 Q_n^2 + \cdots -
\beta E_s + \beta _2 E_s^2 + \cdots ,
\end{equation}
In (\ref{eq5}), $S(E - E_s ,Q - Q_n ) - S(E,Q)$ is the difference
between the entropy before the naked black hole radiates and the
entropy after the naked black hole radiates. That is

\begin{equation}
\label{eq6}
\Delta S = S(E - E_s ,Q - Q_n ) - S(E,Q).
\end{equation}
From Eq.(\ref{eq3}), the energy spectrum of the black hole
radiation is as follows:

\begin{equation}
\label{eq7} \rho _{s,n} \propto e^{\Delta S}.
\end{equation}
According to thermodynamics , $\beta $ should be the inverse of
the temperature.

Normalizing the distribution function, we obtain

\begin{equation}
\label{eq8} \rho _{s,n} = \frac{1}{Z_C }e^{\Delta S} =
\frac{1}{Z_C }e^{ - \beta [E_s + \Phi Q_n ] + \alpha _2 Q_n^2 +
\cdots + \beta _2 E_s^2 + \cdots },
\end{equation}

\noindent where $\Phi = \left( {\frac{\partial E_b }{\partial q}}
\right)_{E_s = 0,Q_n = 0} $ is a static electric potential at the
black hole horizon. $Z_C $ is named as a canonical partition
function. And it is defined as

\begin{equation}
\label{eq9} Z_C = \sum\limits_{E_s,Q_n}{\rho (E - E_s ,Q - Q_n
)e^{\Delta S}} .
\end{equation}

If we adopt the semi-classical method, the partition function in Eq.(\ref{eq9}) can
be rewritten as

\begin{equation}
\label{eq10} Z_C = \int {dE_s dQ_n \rho (E - E_s ,Q - Q_n )e^{ -
\beta [E_s - \Phi Q_n ] + \alpha _2 Q_n^2 + \cdots + \beta _2
E_s^2 + \cdots }} ,
\end{equation}

\noindent where $\rho (E - E_s ,Q - Q_n )$ is the state density
when the energy and the charge are given by $E - E_s $ and $Q -
Q_n $, respectively. In our calculation, when we neglect the
higher-order terms and only keep the first order term, we have

\begin{equation}
\label{eq11}
Z_G = \int {dE_s dQ_n } \rho (E - E_s ,Q - Q_n )e^{ - \beta [E_s - \Phi Q_n
]}.
\end{equation}
When the energy of the radiation particles of the black hole is
$E_s $ and the charge is $Q_n $, the energy of the black hole is
$E_b = E - E_s $, and the charge is $q=Q - Q_n $. For the black
hole, when the energy is $E_b $ and the charge is $q$, the
corresponding state density is $\rho (E - E_s ,Q - Q_n )$.Then

\begin{equation}
\label{eq12}
\rho (E - E_s ,Q - Q_n ) = \exp \left[ {S_{MC} (E - E_s ,Q - Q_n )}
\right].
\end{equation}
The integral in Eq.(\ref{eq11}) can be performed in general by the
saddle point approximation, provided the microcanonical entropy
$S_{MC} (E - E_s ,Q - Q_n )$ can be Taylor-expanded around the
average equilibrium energy $E$ and $Q$

\begin{equation}
\label{eq13} S_{MC} (E - E_s,Q - Q_n) = S_{MC} (E,Q) - \alpha Q_n
+ \alpha _2 Q_n^2 + \cdots - \beta E_s + \beta _2 E_s^2 + \cdots .
\end{equation}
Neglecting the higher-order terms, we get the partition function
given by

\begin{equation}
\label{eq14} Z_C = \int {dE_s dQ_n e^{S_{MC} (E,Q)}e^{2[ - \beta
E_s - \alpha Q_n + \alpha _2 Q_n^2 + \beta _2 E_s^2 ]}} .
\end{equation}
When $\alpha = - \beta \Phi > 0$, the main contribution to the
integral in Eq. (\ref{eq14}) comes from the case when $E_s $ and
$Q_n $ are very small. Letting the integral upper limit be
infinity, we have

\[
Z_C (\beta ,\alpha )
 = e^{S_{MC} (E,Q)}
\left[ {\frac{1}{2}\sqrt {\frac{\pi }{ - 2\beta _2 }} \exp \left(
{\frac{\beta ^2}{ - 2\beta _2 }} \right)\left( {1 - erf\left( {\frac{\beta
}{\sqrt { - 2\beta _2 } }} \right)} \right)} \right]
\times
\]

\begin{equation}
\label{eq15}
\left[ {\frac{1}{2}\sqrt {\frac{\pi }{ - 2\alpha _2 }} \exp \left(
{\frac{\alpha ^2}{ - 2\alpha _2 }} \right)\left( {1 - erf\left(
{\frac{\alpha }{\sqrt { - 2\alpha _2 } }} \right)} \right)} \right].
\end{equation}

\noindent
where

\[
erf(z) = \frac{2}{\sqrt \pi }\int\limits_0^z {e^{ - t^2}} dt
\]

\noindent
is the error function.\vspace{0.5cm}\\
\textbf{\large 3. Canonical entropy of charged black hole}

Using the standard formula from equilibrium statistical mechanics

\begin{equation}
\label{eq16} S_C = \ln {Z_C} - \beta \frac{\partial \ln
{Z_C}}{\partial \beta } - \alpha \frac{\partial \ln
{Z_C}}{\partial \alpha },
\end{equation}

\noindent
it is easy to deduce that the canonical entropy is given in terms of the
microcanonical entropy by

\begin{equation}
\label{eq17}
S_C (E,Q) = S_{MC} (E,Q) + \Delta _S ,
\end{equation}

\noindent
where

\begin{equation}
\label{eq18}
\Delta _S = \ln f(\beta ,\beta _2 ) - \beta \frac{\partial \ln f(\beta
,\beta _2 )}{\partial \beta }
 + \ln f(\alpha ,\alpha _2 ) - \alpha \frac{\partial \ln f(\alpha ,\alpha _2
)}{\partial \alpha },
\end{equation}

\[
f(\beta ,\beta _2 ) = \frac{1}{2}\sqrt {\frac{\pi }{ - 2\beta _2 }} \exp
\left( {\frac{\beta ^2}{ - 2\beta _2 }} \right)
\left[ {1 - erf\left( {\frac{\beta }{\sqrt { - 2\beta _2 } }} \right)}
\right].
\]
Making use of the asymptotic expression of the error function

\begin{equation}
\label{eq19}
erf(z) = 1 - \frac{e^{ - z^2}}{\sqrt \pi z}\left[ {1 + \sum\limits_{k =
1}^\infty {( - 1)^k\frac{(2k - 1)!!}{(2z^2)^k}} } \right],
\quad
\left| z \right| \to \infty ,
\end{equation}

\noindent
we have

\begin{equation}
\label{eq20}
f(\beta ,\beta _2 ) = \frac{1}{2\beta }\left[ {1 + \sum\limits_{k =
1}^\infty {( - 1)^k\frac{(2k - 1)!!}{2^k}\left( {\frac{\sqrt { - 2\beta _2 }
}{\beta }} \right)} ^{2k}} \right].
\end{equation}
In $\Delta _S $, we only consider the logarithm terms. When $\Phi
< 0$, the logarithm terms in $\Delta _S $ are given by

\[
\Delta _S = \ln \left[ {1 + \sum\limits_{k = 1}^\infty {( - 1)^k\frac{(2k -
1)!!}{2^kC^k}} } \right]
\]

\begin{equation}
\label{eq21}
 + \ln \left[ {1 + \sum\limits_{k = 1}^\infty {( - 1)^k\frac{(2k -
1)!!}{2^k}y^k} } \right]
 + 2\ln T - \ln \left| \Phi \right|.
\end{equation}

\noindent where $C$ is the thermal capacity under the condition
that the charge is invariable, $y = \left( {\frac{\partial
}{\partial q}\left( {\frac{1}{\beta \Phi }} \right)} \right)_E $.

In the asymptotic expression of error function given by Eq.
(\ref{eq19}), we take the sum from 1 to $n$ as the approximate
value of the series. When $z$ is a real number, its error does not
exceed the absolute value of the first term that has been
neglected in the series. Therefore, when $C < - 1$ or $C > 1$, the
first term in $\Delta _S $ is not divergent. However, when $ - 1
\le C \le 1$, it is possible that the entropy becomes a complex
number. The fact that the entropy is a complex number implies that
the black hole is not stable in thermodynamics. This is still an
unsolved problem.

When $\left| y \right| < 1$, the second term in $\Delta _S $ is
not divergent. Therefore, we can prove that when the charged black
hole is in the stable state, the relation between the energy and
temperature of the charges satisfies

\begin{equation}
\label{eq22}
\Phi \left( {\frac{\partial Q}{\partial \Phi }} \right)_E < T\left(
{\frac{\partial Q}{\partial T}} \right)_E .
\end{equation}\vspace{0.5cm}\\
\textbf{\large 4. Conclusion and discussion }

For Schwarzschild space-time, when we take the first
approximation, the logarithm correction term to entropy is

\begin{equation}
\label{eq23}
\Delta _S = \ln T = - \frac{1}{2}\ln \frac{A}{4} + const.
\end{equation}
Ref.[17] discussed the correction to the black hole entropy using
the generalized uncertainty relation and derived the following
result.

\begin{equation}
\label{eq24}
S
 = \frac{A}{4} - \frac{\pi \alpha ^2}{4}\ln \left( {\frac{A}{4}} \right) +
\sum\limits_{n = 1}^\infty {C_n } \left( {\frac{A}{4}} \right)^{ - n} +
const.
\end{equation}
Based on Eq.(\ref{eq24}), there is an uncertain factor $\alpha ^2$
in the logarithm term in the correction to the black hole entropy.
However, there is no uncertain factor in our result.

When the correction to the black hole thermodynamic quantities due
to thermal fluctuation is considered, the expression of entropy is
given by [18-21]

\begin{equation}
\label{eq25} S = \ln \rho = S_{MC} - \frac{1}{2}\ln (CT^2) +
\cdots .
\end{equation}
There is a limitation in the above result. Since the thermal
capacity of Schwarzschild black hole is negative. The entropy
given by Eq.(\ref{eq25}) is a complex number. So this relation is
not valid for Schwarzschild black hole. However, when we take a
proper approximation or limit, general four-dimensional curved
space-times can return to Schwarzschild space-times. This implies
that Eq.(\ref{eq25}) is not universal. In our result we only
request the thermal capacity satisfies $C < - 1$ or $C > 1$. A big
Schwarzschild black hole satisfies this condition. According to
this condition, when the energy of Schwarzschild black hole
satisfies $M^2 > 1 / 8\pi $, we can obtain the entropy which is
not divergent. Then we can derive the lowest energy of
Schwarzschild black hole in the universe.

In addition, the research of the black hole entropy is based on the fact
that the black hole has thermal radiation and the radiation spectrum is a
pure thermal spectrum. However, Hawking showed that the radiation spectrum
is a pure thermal spectrum only under the condition that the background of
space-time is invariable. During this radiation process, the information may
get lost. The information loss of the black hole means that the pure quantum
state will disintegrate to a mixed state. This violates the unitarity
principle in quantum mechanics. When we discuss the black hole radiation
using the tunnel effect method, after considering the conversation of energy
and the change of the horizon, we obtain the result that the radiation
spectrum is no longer a strictly pure thermal spectrum. This method avoids
the limitation of Hawking radiation and shows that it is the
self-gravitation which provides the potential barrier of quantum tunnel.

Our discussion is based on the quantum tunnel effect of the black
hole radiation which is independent of that whether static mass of
radiation particle equal to zero. So our discussion is very
reasonable. We provide a way for studying the quantum correction
to Bekenstein-Hawking entropy. Based on our method, we can further
check the string theory and single loop quantum gravity
and determine which one is perfect.\\
ACKNOWLEDGMENT

Zhao Ren thanks Elias Vagenas for useful correspondence and
interesting discussions.This project was supported by the National
Natural Science Foundation of China under Grant No. 10374075 and
the Shanxi Natural Science Foundation of China under Grant No.
2006011012;20021008.

\textbf{REFERENCES}

[1] S. W. Hawking, \textit{Commun. Math. Phys.} \textbf{ 43,}
199(1975 )

[2] M. K. Parikh and F. Wilczek, \textit{Phys. Rev. Lett.}
\textbf{85,} 5042(2000)

[3] E. C. Vagenas, \textit{Phys. Lett}. \textbf{B503}, 399(2001)

[4] E. C. Vagenas, \textit{Mod. Phys. Lett}. \textbf{A17},
609(2002)

[5] E. C. Vagenas, \textit{Phys. Lett.} \textbf{B533}, 302(2002)

[6] A. J. M. Medved, \textit{Class. Quant. Grav}. \textbf{19,
}589(2002)

[7] M. K. Parikh, \textit{Phys. Lett}. \textbf{B546}, 189(2002)

[8] A. J. Medved,\textit{ Phys. Rev}. \textbf{D66}, 124009(2002)

[9] E. C. Vagenas, \textit{Phys. Lett}. \textbf{B559}, 65(2003)

[10] J. Y. Zhang and Z. Zhao, \textit{Phys. Lett}. \textbf{B618},
14(2005); C. Z. Liu, J. Y. Zhang and Z. Zhao, \textit{Phys. Lett.}
\textbf{B639}, 670(2006)

[11] J. Y. Zhang and Z. Zhao, \textit{Nucl. Phys}. \textbf{B725},
173 (2005);\textbf{ JHEP10}, 055(2005)

[12] M. Angheben, M. Nadalini, L. Vanzo and S. Zerbini,
\textbf{JHEP05}, 014(2005)

[13] A. J. M. Medved and E. C. Vagenas, \textit{Mod. Phys.
Lett}.\textbf{A20}, 2449(2005)

[14] A. J. M. Medved and E. C. Vagenas, \textit{Mod. Phys. Lett.}
\textbf{A20}, 1723(2005)

[15] M. Arzan, A. J. M. Medved and E. C. Vagenas, \textbf{JHEP09},
037(2005)

[16] M. Arzano, \textit{Mod. Phys. Lett.} \textbf{A21}, 41(2006)

[17] A. J. M. Medved and E. C. Vagenas, \textit{Phys. Rev.}
\textbf{D70}, 124021(2004)

[18] M. R. Setare, \textit{Phys. Lett.} \textbf{B573}, 173(2003)

[19] M. Cavaglia and A. Fabbri, \textit{Phys. Rev.} \textbf{D65},
044012 (2002)

[20] G. Gour and A.J. M. Medved, \textit{Class. Quant. Grav.}
\textbf{20}, 3307(2003)

[21] M. R. Setare, \textit{Eur. Phys.} J. C \textbf{33}, 555(2004)

\end{document}